\documentclass[reqno]{amsart}

\newcommand{\Rm}{\mathbb{R}}

\newcommand{\Sm}{\mathbb{S}}
\newcommand{\be}{\begin{equation}}
\newcommand{\ee}{\end{equation}}
\newcommand{\va}{\varphi}

\newcommand{\pp}{\partial}

\newcommand{\ket}[1]{\left|#1\right\rangle}
\newcommand{\bra}[1]{\left\langle#1\right|}
\newcommand{\braket}[2]{\left\langle#1\middle|#2\right\rangle}

\usepackage{amsmath}
\usepackage[foot]{amsaddr}
\usepackage{amsthm,amssymb,mathrsfs}
\usepackage{graphicx}
\usepackage{color}

\newtheorem{thm}{Theorem}[section]
\newtheorem{lem}[thm]{Lemma}

\theoremstyle{remark}\newtheorem{rmk}[thm]{Remark}

\title[]{Chandrasekhar polynomials -- A brief review}

\author{Manabu Machida}
\email{machida@hama-med.ac.jp}
\address{Institute for Medical Photonics Research,
Hamamatsu University School of Medicine,
Hamamatsu 431-3192, Japan}

\begin{document}

\begin{abstract}
A review on the Chandrasekhar polynomials is given. The polynomials often appear in transport theory. The relation to the method of rotated reference frames for the three-dimensional radiative transport equation is clarified.
\end{abstract}

\maketitle

\section{Introduction}
\label{intro}
The Chandrasekhar polynomials play an important role in one-dimensional transport theory (see \cite{Ganapol14,Kuscer-McCormick91} and references therein). Recently, the appearance of the polynomials has been recognized even for the three-dimensional radiative transport equation \cite{Machida14,Machida15}.

We begin with the one-dimensional transport equation. Let $\mu_t,\mu_s$ be constants such that $\mu_t>\mu_s\ge0$. Let $\mu$ be the third component of vector $\theta\in\Sm^2$, i.e., $\mu$ is the cosine of the polar angle of $\theta$. Let $\Omega$ be an interval on the real axis. We write the transport equation as
\[
\left(\mu\frac{\pp}{\pp z}+\mu_t\right)I(z,\theta)=
\mu_s\int_{\Sm^2}p(\theta,\theta')I(z,\theta')\,d\theta',\quad
(z,\theta)\in\Omega\times\Sm^2.
\]
The solution $I(z,\theta)$ will be uniquely determined if suitable boundary conditions are imposed. We assume that the scattering phase function $p(\theta,\theta')$ is given by
\[
p(\theta,\theta')=\frac{1}{4\pi}\sum_{l=0}^L\beta_lP_l(\theta\cdot\theta')=
\sum_{l=0}^L\sum_{m=-l}^l\frac{\beta_l}{2l+1}Y_{lm}(\theta)Y_{lm}^*(\theta'),
\]
where $P_l$ are Legendre polynomials, $Y_{lm}$ are spherical harmonics, and the symbol $*$ means complex conjugate. The coefficient $\beta_0=1$ and for $1\le l\le L$, $|\beta_l|<2l+1$. Using associated Legendre polynomials $P_l^m(\mu)$, spherical harmonics are given by
\[
Y_{lm}(\theta)=\sqrt{\frac{2l+1}{4\pi}\frac{(l-m)!}{(l+m)!}}P_l^m(\mu)e^{im\va},
\]
where $\va\in[0,2\pi)$ is the azimuthal angle of $\theta$. We note that this $p(\theta,\theta')$ implies scatterers are spherically symmetric. In optics, coefficients $\beta_l$ are often given by $\beta_l=(2l+1){\rm g}^l$ with the anisotropy factor ${\rm g}\in(-1,1)$ \cite{Bigio-Fantini}.

By changing the spatial variable as $x=\mu_tz$, we can rewrite the transport equation as
\[
\left(\mu\frac{\pp}{\pp x}+1\right)\psi(x,\theta)=
\varpi\int_{\Sm^2}p(\theta,\theta')\psi(x,\theta')\,d\theta',\quad
(x,\theta)\in\Omega\times\Sm^2,
\]
where $\varpi=\mu_s/\mu_t\in[0,1)$ is called the albedo for single scattering and $\psi(x,\theta)=I(x/\mu_t,\theta)$. Chandrasekhar's polynomials appear when the solution $\psi(x,\theta)$ to the homogeneous equation is sought assuming the form
\[
\psi(x,\theta)=\sum_{l=0}^{\infty}\sum_{m=-l}^lf_{lm}(\nu)Y_{lm}(\theta)e^{-x/\nu},
\]
where $\nu\in\Rm$ is a parameter and $f_{lm}(\nu)$ are coefficients which will be later related to Chandrasekhar's polynomials. See \cite{Barichello-Siewert98} for the equivalence between the method of discrete ordinates and the spherical-harmonic expansion.

Let us introduce $h_l$ as
\[
h_l=\left\{\begin{aligned}
2l+1-\varpi\beta_l,&\quad 0\le l\le L,
\\
2l+1,&\quad l\ge L+1.
\end{aligned}\right.
\]
We note the relation
\[
\int_{-1}^1\mu P_l^m(\mu)P_{l'}^m(\mu)\,d\mu=
\frac{2}{4(l+1)^2-1}\frac{(l+1+m)!}{(l-m)!}\delta_{l+1,l'}+
\frac{2}{4l^2-1}\frac{(l+m)!}{(l-1-m)!}\delta_{l-1,l'}.
\]
By substituting the assumed form of $\psi(x,\theta)$ into the homogeneous transport equation, we obtain
\[
\mu\sum_{l'=0}^{\infty}\sum_{m'=-l'}^{l'}f_{l'm'}(\nu)Y_{l'm'}(\theta)
-\nu\sum_{l'=0}^{\infty}\sum_{m'=-l'}^{l'}\frac{h_{l'}}{2l'+1}f_{l'm'}(\nu)Y_{l'm'}(\theta)=0.
\]
Then by multiplying $Y_{lm}^*(\theta)$ and integrating over $\theta$, we obtain
\[
\sum_{l'=|m|}^{\infty}\sqrt{\frac{l^2-m^2}{4l^2-1}}\delta_{l-1,l'}f_{l'm}(\nu)-\frac{\nu h_l}{2l+1}f_{lm}(\nu)+\sum_{l'=|m|}^{\infty}\sqrt{\frac{(l+1)^2-m^2}{4(l+1)^2-1}}\delta_{l+1,l'}f_{l'm}(\nu)=0.
\]
Let us define $f_{lm}(\nu)=0$ for $l<|m|$. If we multiply $\sqrt{2l+1}$ in the above equation, we obtain
\[
\sqrt{\frac{l^2-m^2}{2l-1}}f_{l-1,m}(\nu)-\frac{\nu h_l}{\sqrt{2l+1}}f_{lm}(\nu)+\sqrt{\frac{(l+1)^2-m^2}{2(l+1)+1}}f_{l+1,m}(\nu)=0.
\]

\section{Chandrasekhar polynomials}
\label{polyg}

Let $x\in\Rm$. Chandrasekhar introduced polynomials $G_l^m(x)$ which satisfy the following three-term recurrence relation \cite{Benassi-etal84,Chandrasekhar}.
\[
(l+m)G_{l-1}^m(x)-h_lxG_l^m(x)+(l-m+1)G_{l+1}^m(x)=0,\quad l\ge m\ge 0,
\]
with $G_m^m(x)=(2m-1)!!$. See \cite{Inonu70} for the case $m=0$.

Then the normalized Chandrasekhar polynomials $g_l^m(x)$ ($l\ge|m|$) were introduced \cite{Garcia-Siewert89,Garcia-Siewert90}. By setting \cite{Siewert-McCormick97}
\[
g_l^m(x)=\sqrt{\frac{(l-m)!}{(l+m)!}}G_l^m(x),
\]
we see that $g_l^m$ satisfy the following three-term recurrence relation.
\[
\sqrt{l^2-m^2}g_{l-1}^m(x)-h_lxg_l^m(x)+\sqrt{(l+1)^2-m^2}g_{l+1}^m(x)=0,
\quad l\ge|m|.
\]
Indeed, the three-term recurrence relation for $f_{lm}$ is recovered if we put $g_l^m=f_{lm}/\sqrt{2l+1}$. We set the initial term as
\[
g_m^m(x)=\frac{(2m-1)!!}{\sqrt{(2m)!}}=\frac{\sqrt{(2m)!}}{2^mm!},
\quad m\ge0.
\]
Moreover, $g_l^{-m}(x)$ and $g_l^m(-x)$ are related to $g_l^m(x)$ as
\[
g_l^{-m}(x)=(-1)^mg_l^m(x),\quad
g_l^m(-x)=(-1)^{l+m}g_l^m(x).
\]

\section{Eigenproblem}

To avoid tedious calculations, in this section we assume $m$ is nonnegative: $m=0,1,\dots$. It is straightforward to extend results below to the case of negative $m$. It is also possible to write $p(\theta,\theta')$ only with $m\ge0$ making use of the formula $P_l^{-m}(\mu)=(-1)^m\frac{(l-m)!}{(l+m)!}P_l^m(\mu)$. Let us introduce
\[
\sigma_l=\frac{\mu_th_l}{2l+1},
\]
and
\[
y_l^m(x)=\sqrt{(2l+1)\sigma_l}g_l^m(x).
\]
Using the new notation, the three-term recurrence relation for $g_l^m$ becomes
\[
b_l(m)y_{l-1}^m(x)-\frac{x}{\mu_t}y_l^m(x)+b_{l+1}(m)y_{l+1}^m(x)=0,
\]
where
\[
b_l(m)=\sqrt{\frac{l^2-m^2}{(4l^2-1)\sigma_l\sigma_{l-1}}}.
\]

By imposing the truncation condition
\[
g_{M+1}^m(\xi)=0,\qquad M=l_{\rm max}+m\quad\mbox{or}\quad l_{\rm max},
\]
where $M$ determines the highest degree of $P_l^m$ used to express $\psi(x,\theta)$, we arrive at the eigenproblem
\[
B(m)Y_{\xi}(m)=\frac{\xi}{\mu_t}Y_{\xi}(m),
\]
where $Y_{\xi}(m)=(y_m^m(\xi),y_{m+1}^m(\xi),\dots,y_M^m(\xi))^T$. The tridiagonal matrix $B(m)$ is given by \cite{Markel04,Panasyuk-etal06}
\[
\{B(m)\}_{ll'}=b_l(m)\delta_{l',l-1}+b_{l'}(m)\delta_{l',l+1}.
\]
In the method of rotated reference frames \cite{Markel04,Panasyuk-etal06}, eigenmodes are labeled by eigenvalues of $B(m)$. The number of rows and columns of $B(m)$ is $l_{\rm max}+1$ when $M=l_{\rm max}+m$ and is $l_{\rm max}-m+1$ when $M=l_{\rm max}$. Since $B(m)$ is a symmetric tridiagonal matrix with nonzero off-diagonal elements, its eigenvalues are distinct. Also if $\xi/\mu_t$ is an eigenvalue for $y_l^m(\xi)$, then $-\xi/\mu_t$ is another eigenvalue and $y_l^m(-\xi)=(-1)^ly_l^m(\xi)$ \cite{Markel04}. Essentially the same tridiagonal matrix $W$ was introduced in \cite{Siewert-McCormick97}. Elements of $W$ are given by $\{W\}_{ll'}=w_l(m)\delta_{l',l-1}+w_{l'}(m)\delta_{l',l+1}$, where $w_l(m)=\sqrt{(l^2-m^2)/(h_lh_{l-1})}$. Let $\xi_j/\mu_t$ ($j=1,\dots,l_{\rm max}+1$) denote eigenvalues of $B(m)$. We note that $\{\xi_j\}$ are eigenvalues of $W$.

For simplicity, hereafter, we suppose $M=l_{\rm max}+m$ and $l_{\rm max}\ge1$ is an odd integer. There are $(l_{\rm max}+1)/2$ positive eigenvalues and $(l_{\rm max}+1)/2$ negative eigenvalues for each $m$. Then we can write eigenvalues as
\[
\xi_1>\xi_2>\cdots>\xi_{\frac{l_{\rm max}+1}{2}}>0>\xi_{\frac{l_{\rm max}+1}{2}+1}>\cdots>\xi_{l_{\rm max}+1},
\]
and $\xi_{l_{\rm max}+2-j}=-\xi_j$ ($j=1,\dots,(l_{\rm max}+1)/2$).

The following lemmas hold.

\begin{lem}[Orthogonality \cite{McCormick95,Siewert-McCormick97}]
\label{lem:ortho}
We have
\[
\frac{1}{Z_j}\sum_{l=m}^{l_{\rm max}+m}y_l^m(\xi_i)y_l^m(\xi_j)=\delta_{ij},
\quad i,j=1,2,\dots,l_{\rm max}+1,
\]
where $Z_j=\sum_{l=m}^{l_{\rm max}+m}[y_l^m(\xi_j)]^2$.
\end{lem}

\begin{proof}
Eigenvectors corresponding to two distinct eigenvalues of a symmetric real matrix are orthogonal.
\end{proof}

\begin{lem}[Completeness \cite{McCormick95,Siewert-McCormick97}]
\label{lem:compl}
We have
\[
\sum_{j=1}^{l_{\rm max}+1}\frac{1}{Z_j}y_l^m(\xi_j)y_{l'}^m(\xi_j)=\delta_{ll'}
,\quad l,l'=m,m+1,\dots,m+l_{\rm max},
\]
where $Z_j=\sum_{l=m}^{l_{\rm max}+m}[y_l^m(\xi_j)]^2$.
\end{lem}

\begin{proof}
Let us introduce vectors $X_j=Y_{\xi_j}(m)/\sqrt{Z_j}$ and matrix $X=(X_1,\dots,X_{l_{\rm max}+1})$. Then $X$ is an orthogonal matrix: $X^{-1}=X^T$. Next we introduce matrices $Z=\mathop{\mathrm{diag}}(\sqrt{Z_1},\dots,\sqrt{Z_{l_{\rm max}+1}})$ and $Y=(Y_{\xi_1}(m),\dots,Y_{\xi_{l_{\rm max}+1}}(m))$. The matrix $Y$ is expressed as $Y=XZ$.

Let us consider
\[
\sum_{j=1}^{l_{\rm max}+1}D_jy_{l'}^m(\xi_j)=\delta_{ll'},\quad
l,l'=m,m+1,\dots,m+l_{\rm max}.
\]
To find $D_j$ ($j=1,\dots,l_{\rm max}+1$), we introduce vectors $D=(D_1,\dots,D_{l_{\rm max}+1})^T$ and $F=(\delta_{m,l},\delta_{m+1,l},\dots,\delta_{m+l_{\rm max},l})^T$, and write the relation as $YD=F$. Since $D=Z^{-1}X^TF$, we obtain
\[
D_j=\frac{1}{Z_j}y_l^m(\xi_j).
\]
This completes the proof.
\end{proof}

\begin{rmk}
If we define $\ket{y_{\xi_j}(m)}=Y_{\xi_j}(m)/\sqrt{Z_j}$ and $\braket{l}{y_{\xi_j}(m)}=y_l^m(\xi_j)/\sqrt{Z_j}$, then the orthogonality and completeness in Lemma \ref{lem:ortho} and Lemma \ref{lem:compl} are equivalently expressed as
\[
\braket{y_{\xi_i}(m)}{y_{\xi_j}(m)}=\delta_{ij},\quad
\sum_{j=1}^{l_{\rm max}+1}\ket{y_{\xi_j}(m)}\bra{y_{\xi_j}(m)}=1.
\]
\end{rmk}

\section{Concluding remarks}
\label{concl}

In this paper, we focused on the case $\mu_a>0$, i.e., $\varpi<1$. It is possible to consider the conservative (nonabsorbing) case $\mu_a=0$ but it must be done separately. When $\varpi=1$, $\sigma_0=0$ and the element $b_1(0)$ becomes infinity. In this case, we need to remove the top left part of $B(0)$.

In application, the numerical evaluation of the Chandrasekhar polynomials is important. Various numerical techniques have been developed \cite{Ganapol14,Garcia-Siewert89,Garcia-Siewert90}.

\section*{Acknowledgements}
The author acknowledges support from JSPS KAKENHI Grant No.~17K05572, 18K03438.

%\newpage                                                            
%\setcounter{section}{1}
%\appendix

%\section*{References}
%\bibliographystyle{plain}
%\bibliography{myref}

\end{document}